# De Broglie Wavelength of a Nonlocal Four-Photon

Philip Walther*, Jian-Wei Pan†, Markus Aspelmeyer*, Rupert Ursin*, Sara Gasparoni* & Anton Zeilinger*‡

***** *Institut für Experimentalphysik, Universität Wien, Boltzmanngasse 5, 1090 Wien, Austria*

† *Physikalisches Institut, Universität Heidelberg, Philosophenweg 12, D-69120 Heidelberg, Germany*

‡ *Institut für Quantenoptik und Quanteninformation, Österreichische Akademie der Wissenschaften*

**Superposition is one of the most distinct features of quantum theory and has been demonstrated in numerous realizations of Young's classical double-slit interference experiment and its analogues[1-5]. However, quantum entanglement[6] - a significant coherent superposition in multiparticle systems - yields phenomena that are much richer and more interesting than anything that can be seen in a one-particle system[7,8]. Among them, one important type of multi-particle experiments uses path-entangled number-states, which exhibit pure higher-order interference and allow novel applications in metrology and imaging[9] such as quantum interferometry and spectroscopy with phase sensitivity at the Heisenberg limit[10-12] or quantum lithography beyond the classical diffraction limit[13]. Up to now, in optical implementations of such schemes lower-order interference effects would always decrease the overall performance at higher particle numbers. They have thus been limited to two photons[14]. We overcome this limitation and demonstrate a linear-optics-based four-photon interferometer. Observation of a four-particle mode-entangled state is confirmed by interference fringes with a periodicity of one quarter of the single-photon wavelength. This scheme can readily be extended to arbitrary photon numbers and thus represents an important step towards realizable applications with entanglement-enhanced performance.**

To see the origin of multiparticle interference more clearly, consider first a simple analogue to Young's double slit experiment, i.e. a Mach-Zehnder (MZ) interferometer (Figure 1). There, single-photon interference occurs due to the spatial separation of two modes of propagation a1 and b1 for a single particle entering the interferometer at the first beamsplitter. Variation of the path length Δx induces a phase shift Δφ and gives thus rise to detection probabilities $P_{a2} \propto 1 + \cos \Delta j$ and $P_{b2} \propto 1 - \cos \Delta j$ in each of the two output modes a2 and b2 behind the exit beamsplitter. Two-photon interference occurs when a1 and b1 are the modes of propagation for a state of two indistinguishable photons, i.e. a biphoton state $|\Psi\rangle = \frac{1}{\sqrt{2}}\left(|2\rangle_{a1}|0\rangle_{b1} + e^{i2\Delta j}|0\rangle_{a1}|2\rangle_{b1}\right)$. This is a superposition where either two photons are in mode a1 and none are in mode b1, or, vice versa, no photons are in mode a1 and two photons propagate within mode b1. This represents a path-entangled two-photon state, which exhibits pure two-particle interference at the output beamsplitter. Note that the path length variation Δx acts on both photons and gives thus rise to a doubled phase shift compared to the single photon case. Because of the unavailability of detectors which are able to distinguish between $N$ and $N+1$ photons, coincidence measurements of spatially separated photons are required to observe multi-photon states. This can be achieved, though only probabilistically, by adding a beamsplitter in each of the spatial output modes of the interferometer. The probability to find two photons in either mode a2 or b2 then oscillates with $P_{a2,a2} \propto 1+\cos(2\Delta j)$ and $P_{b2,b2} \propto 1-\cos(2\Delta j)$, respectively, while the single-photon detection probabilities $P_{a2}$ and $P_{b2}$ remain constant.

In the generalized case of an $N$-particle interferometer, the $N$ photons will be in a superposition of being in either mode a1 or b1, resulting in

$$\left|\Psi\right\rangle = \frac{1}{\sqrt{2}}\left(\left|N\right\rangle_{a1}\left|0\right\rangle_{b1} + e^{iN\Delta j}\left|0\right\rangle_{a1}\left|N\right\rangle_{b1}\right). \tag{1}$$

In other words, the paths are entangled in photon number. Here $\left|N\right\rangle_{a1}$ (or $\left|N\right\rangle_{b1}$) indicates the $N$-particle Fock state in spatial mode a1 (or b1), respectively, and $N=0$ represents an empty mode. The phase modulation $N\Delta j$ increases linearly with the particle number $N$, which is the origin of all entanglement-enhanced interferometric schemes. In particular, the $N$-photon detection probability in each of the interferometer outputs would vary as $P_N \propto 1+\cos(N\Delta j)$. It has therefore been suggested to attribute an effective de Broglie wavelength ?/$N$ to the quantum state. This resembles the case of a heavy massive molecule consisting of $N$ atoms; though here the particles are in no way bound to each other[15].

In order to best benefit from such entanglement-enhanced interferometric techniques it is desirable to experimentally achieve a high photon number $N$ for states of the form (1). The special case of $N = 2$ was realized both in the original Young's double-slit geometry by using collinear production of biphoton states via parametric down conversion[16] and in Mach-Zehnder configuration by using two-photon interference to suppress unwanted single-photon contributions[17,18]. It is commonly believed that the realization for states with $N > 2$ requires the use of non-linear gates[19] or $N$ additional "ancilla" detectors with single-photon resolution[20]. Unfortunately, each of these schemes is not feasible with current technologies. We demonstrate how to overcome this limitation giving a specific example of pure four-photon interferometry.

Our proposal is based on separating photon pairs into different pairs of modes and utilizing two-particle-interferometry rather than distinguishing photon numbers or employing nonlinear beamsplitters. To achieve this goal, we exploit type-II spontaneous parametric down-conversion (SPDC)[21]. An ultra-violet pulse passes through a beta-barium-borate (BBO) crystal, probabilistically emitting pairs of energy-degenerate polarization-entangled photons into the spatial modes a1 and a2 (Figure 2). The UV pump beam is reflected back at a mirror and can thus also emit entangled photon pairs into the spatial modes b1 and b2 (Figure 2). The setup is aligned to generate the following maximally entangled biphoton state

$$\left|\Phi^+\right\rangle = \frac{1}{\sqrt{2}}\left(\left|H\right\rangle_1\left|H\right\rangle_2 + \left|V\right\rangle_1\left|V\right\rangle_2\right) \tag{3}$$

for each of the pairs emitted into the pairs of modes a1-a2 and b1-b2, respectively. Here $H$ (or $V$) indicates horizontal (or vertical) polarization of the photon.

We first consider the case where only one pair of entangled photons is emitted on a double pass of the UV pulse through the crystal. There are two probability amplitudes which will contribute to the emerging two-photon state, i.e. the pair is emitted either into the pair of modes a1-a2 or into the pair of modes b1-b2. We then coherently combine the two pairs of modes at the two polarizing beamsplitters (PBS). Since the PBS transmits horizontally polarized light and reflects vertically polarized light, conditional on detecting one photon in each of the output ports a3 and a4 the biphoton state will be

$$\left|\Phi\right\rangle_{a3a4} = \frac{1}{\sqrt{2}}\left(\left|\mathrm{H}\right\rangle_{a3}\left|\mathrm{H}\right\rangle_{a4} + e^{i2\Delta\varphi}\left|\mathrm{V}\right\rangle_{a3}\left|\mathrm{V}\right\rangle_{a4}\right) \quad (4)$$

where again $\Delta\varphi$ is the phase modulation of a single-photon[17, 22]. The phase $\Delta\varphi$ is proportional to the position of the pump mirror PM, where $\Delta x$ is the UV path-length change. Two-photon interference fringes may now be observed by performing a projection measurement in the modes a3 and a4 in the linear polarization basis $\left|\pm\right\rangle = \left(1/\sqrt{2}\right)\left(\left|H\right\rangle \pm \left|V\right\rangle\right)$. Specifically, the probability of detecting a two-fold coincidence $\left|+\right\rangle_{a3}\left|-\right\rangle_{a4}$ is proportional to $P_{a3,a4} \propto 1 - \cos\left(2\Delta\varphi\right)$. These correlations are already a signature of non-locality[23,24] of the two-photon state (Figure 3b).

Let us now explain how our scheme can be generalized to four photons and even higher photon numbers. Consider the case where two pairs of photons are emitted on a double pass of the pump beam through the crystal. There are two possibilities which will contribute to an overall four-photon state, i.e. either by double-pair emission on one or the other side, where two photon pairs are emitted into the same pair of modes a1-a2 or b1-b2, respectively, or one pair of photons is emitted simultaneously into each of the modes a1-a2 and b1-b2.

We first study the double-pair emission case where both pairs propagate within the same mode pair. A four-fold coincidence after the two PBS, i.e. detection of a single photon in each of the output ports a3, a4, b3 and b4, will either result from a $|H\rangle_{a3}|H\rangle_{a4}|V\rangle_{b3}|V\rangle_{b4}$ contribution, if the two photon pairs are both in a1-a2, or from a $|V\rangle_{a3}|V\rangle_{a4}|H\rangle_{b3}|H\rangle_{b4}$ contribution, if the two photon pairs are both in b1-b2. Temporal overlapping of both pairs of modes at the two PBS results in $|H\rangle_{a3}|H\rangle_{a4}|V\rangle_{b3}|V\rangle_{b4} + e^{i4\Delta j}|V\rangle_{a3}|V\rangle_{a4}|H\rangle_{b3}|H\rangle_{b4}$, a coherent superposition of forward and backward emission at the same time, where all the four backward emitted photons are phase shifted by the pump mirror PM. Introducing a path difference of Dx and further performing a polarization measurement in the $|\pm\rangle$ basis to achieve indistinguishability results in interference fringes with one quarter of the single-photon wavelength.

However, with the same probability as the double-pair emission, one photon pair is emitted into each of the mode pairs a1-a2 and b1-b2 by one pulse. That second case of one pair being emitted forwards and one backwards will also result in four-fold coincidences, either from a $|H\rangle_{a3}|H\rangle_{a4}|H\rangle_{b3}|H\rangle_{b4}$ or $|V\rangle_{a3}|V\rangle_{a4}|V\rangle_{b3}|V\rangle_{b4}$ contribution, where all the photons have the same polarization. This coherent superposition $e^{i2\Delta j}\left(|H\rangle_{a3}|H\rangle_{a4}|H\rangle_{b3}|H\rangle_{b4} + |V\rangle_{a3}|V\rangle_{a4}|V\rangle_{b3}|V\rangle_{b4}\right)$ has an overall phase and thus a fixed relative phase[25] which is independent of the position of the pump mirror. Note that this overall phase is halved compared to the double-pair emission case. This is due to the fact that in this case in each contribution only two photons are affected by the pump mirror. Therefore the complete (for simplicity un-normalized) four-photon state behind the PBS can be written as:

$$\begin{aligned}&|H\rangle_{a3}|H\rangle_{a4}|V\rangle_{b3}|V\rangle_{b4} + e^{i4\Delta j}|V\rangle_{a3}|V\rangle_{a4}|H\rangle_{b3}|H\rangle_{b4} + \\ &e^{i2\Delta j}\left(|H\rangle_{a3}|H\rangle_{a4}|H\rangle_{b3}|H\rangle_{b4} + |V\rangle_{a3}|V\rangle_{a4}|V\rangle_{b3}|V\rangle_{b4}\right)\end{aligned} \tag{5}$$

For observing undisturbed four photon interference the last two contributions have to be erased. This can be achieved by performing a proper projection measurement of the four output modes a3, b3, a4 and b4 into the $|\pm\rangle$ bases; then the number of $|+\rangle$ projections is different from the number of $\left|-\right\rangle$ projections, say $\left|+\right\rangle_{a3}\left|-\right\rangle_{a4}\left|+\right\rangle_{b3}\left|+\right\rangle_{b4}$. The overall four-photon amplitude originating from one photon in each mode then becomes $e^{i2\Delta j}\left(\left|+\right\rangle_{a3}\left|-\right\rangle_{a4}\left|+\right\rangle_{b3}\left|+\right\rangle_{b4}-\left|+\right\rangle_{a3}\left|-\right\rangle_{a4}\left|+\right\rangle_{b3}\left|+\right\rangle_{b4}\right)$ and thus vanishes due to the fixed phase relation. This is the four-photon equivalent to a Hong-Ou-Mandel interference[26] of two photons arriving at the beamsplitter. Thus the four-photon detection probability in the spatially separated output modes a3, a4, b3 and b4 oscillates like $P_{a3,a4,b3,b4} \propto 1+\cos(4\Delta j)$, i.e. this projection allows the observation of unperturbed interference of a four-photon state.

Figure 3 compares this pure four-photon interference effect (Figure 3c) with the well-known two-photon interference (Figure 3b) and the single-photon interference (Figure 3a) as were obtained with the same setup. Fits to the data reveal a reduction of the oscillation wavelength from 823±46 nm for the single-photon case over 395± 16 nm for the two-photon case and 194±9 nm for the four-photon case. The deviation is within experimental error given by the thermal long-term stability of our interferometric setup. This demonstrates that the phase modulation of the mirror PM is applied to all the spatially separated four photons simultaneously. Consequently, one has to treat the four-photon state (5) as one object of the form $\left|y\right\rangle=\frac{1}{\sqrt{2}}\left(\left|4\right\rangle_{a1,a2}\left|0\right\rangle_{b1,b2}+e^{i4\Delta j}\left|0\right\rangle_{a1,a2}\left|4\right\rangle_{b1,b2}\right)$, which is similar to a so-called "noon"-state[27]. Our four-photon state has the additional interesting property that it is nonlocal. It is a superposition of four photons either in mode a1 and a2 or b1 and b2. The de-Broglie wavelength feature is then realized by a joint nonlocal measurement on the two photons in a3 and b3 on one side together with the photons a4 and b4 on the other.

In contrast, any projection measurement different from the above will result in an equal contribution of all four-photon terms to the interference pattern, including the unwanted contributions of one photon per spatial mode, as has been observed before[28]. For example, no pure four-fold wavelength reduction can be attained by projecting the four-photon state of (5) onto $|+\rangle_{a3}|+\rangle_{a4}|+\rangle_{b3}|+\rangle_{b4}$. There, the four-photon detection probability oscillates like $P_{a3,a4,b3,b4} \propto 1+\cos^2(2\Delta j) = \frac{3}{2} + 2\cos(2\Delta j) + \frac{1}{2}\cos(4\Delta j)$ , which is confirmed by comparing the simultaneously measured two- and four-photon coincidences of Figure 4.

The employed method allows the generation of four-photon states and their subsequent utilization in pure four-particle interferometry. The result clearly confirms the theoretical expectation that the de-Broglie wavelength of a four-photon state is one fourth of a single photon, thus leading to the general rule $l(N) = l(1)/N$ . This overcomes state-of-the-art two-particle interferometry and opens new possibilities in quantum metrology and in quantum imaging applications, which might be a potentially useful tool for nano-technology[27]. It is important to note that, in principle, this scheme can be extended to higher particle numbers if more spatial modes are involved. The actual limitation due to low count rates might eventually be overcome with the next generation of entangled photon sources.

**Acknowledgements**: We thank C. Brukner and K. Resch for discussions. This work was supported by the Austrian Science Foundation (FWF), project number SFB 015 P06, by the European Commission, contract no. IST-2001-38864, RAMBOQ, and by the Alexander von Humboldt-Foundation.

**Figure1** A two-mode Mach-Zehnder interferometer. The phase is changed by varying the path length via the position of a mirror. Single-photon interference occurs due to the spatial separation of two possible modes of propagation a1 and b1 for a single particle entering the interferometer at the first beamsplitter (BS). Two-photon interference can be achieved when a1 and b1 are the two possible modes of propagation for a biphoton state.

**Figure 2** In our experiment the required four-photon state is produced by type-II spontaneous parametric down-conversion (SPDC). A 200 fs pulse at a central UV-wavelength of 395 nm and at a repetition rate of 76 MHz passes through a beta-barium-borate (BBO) crystal probabilistically emitting pairs of energy-degenerate polarization-entangled photons at 790 nm into the spatial modes a1 and a2. The UV pump beam is reflected back at a mirror and might thus emit a second pair into the spatial modes b1 and b2. The probability of single-pair creation is on the order of p (in our setup $10^{-2}$ – $10^{-3}$), while the probability to create two pairs is proportional to $p^2$. 3nm bandwidth filters (F) and coupling into single-mode fibres in front of each detector enables good temporal and spatial overlap of the photon-wavepackets at the polarizing beamsplitters (PBS). The UV-pump is reflected by the pump mirror PM, which is mounted on a computer-controlled translation stage. By scanning the position of PM with a step size of 1 µm and performing fine adjustment of the position of M, we achieved the temporal overlap of modes a1 and b1, and of modes a2 and b2. An additional piezo translation stage is used to move the pump mirror PM and to perform a change of the phase between four photons emitted into modes a1 and a2 relative to the four photons emitted into b1 and b2. The detection of the spatially separated 4-photon coincidences behind a 45° polarizer (Pol) while varying the position of PM leads to the observed interference fringes.

**Figure 3** Experimental demonstration of pure one-, two- and four-photon interference. The two- and four- photon interference is recorded simultaneously, while for the one-photon interferometry the pulsed laser has been switched from mode-locking to continuous-wave (cw) mode. **a** Single photon rate in mode a3 after performing a projection measurement in the linear polarization basis $|\pm\rangle = (1/\sqrt{2})(|H\rangle \pm |V\rangle)$. For this interference pattern, the pump laser is used in the CW mode at 790nm (instead of the mode-locked frequency-doubled mode at 395nm). A Mach-Zehnder configuration for modes a1-b1 arises for light scattered from the BBO-crystal when passing through the crystal. By moving the pump mirror PM interference fringes appear for single photons with a central wavelength of 790nm which corresponds to the down-converted photons. Note that, due to the back reflection of the pump beam, the change in the optical path is twice as large as in the position of the pump mirror. **b** The two-photon coincidence rate corresponding to the detection in mode a3 and a4 after projecting onto $|+\rangle_{a3}|-\rangle_{a4}$. **c** Performing a projection onto $|+\rangle_{a3}|-\rangle_{a4}|+\rangle_{b3}|+\rangle_{b4}$ results in pure four-photon interference due to projection onto the (non-local) path-entangled four-photon state $|y\rangle = \frac{1}{\sqrt{2}}\left(|4\rangle_{a1,a2}|0\rangle_{b1,b2} + e^{i4\Delta j}|0\rangle_{a1,a2}|4\rangle_{b1,b2}\right)$.

**Figure 4** Two- and four-photon interference without proper post-selection. **a** Two-photon interference after passing two 45° polarizer, projecting onto $|+\rangle_{a3}|+\rangle_{a4}$ in the mode a3 and a4. **b** Simultaneously measured four-photon coincidences after projecting onto the state $|+\rangle_{a3}|+\rangle_{a4}|+\rangle_{b3}|+\rangle_{b4}$. This leads to additional, unwanted two-photon interference terms resulting in the maximally entangled state $|\Psi_2\rangle \propto |4\rangle_{a1,a2}|0\rangle_{b1,b2} + e^{i4\Delta j}|0\rangle_{a1,a2}|4\rangle_{b1,b2} + 2e^{i2\Delta j}|2\rangle_{a1,a2}|2\rangle_{b1,b2}$. Again, the effective change in the optical path is twice the movement of the pump mirror PM.

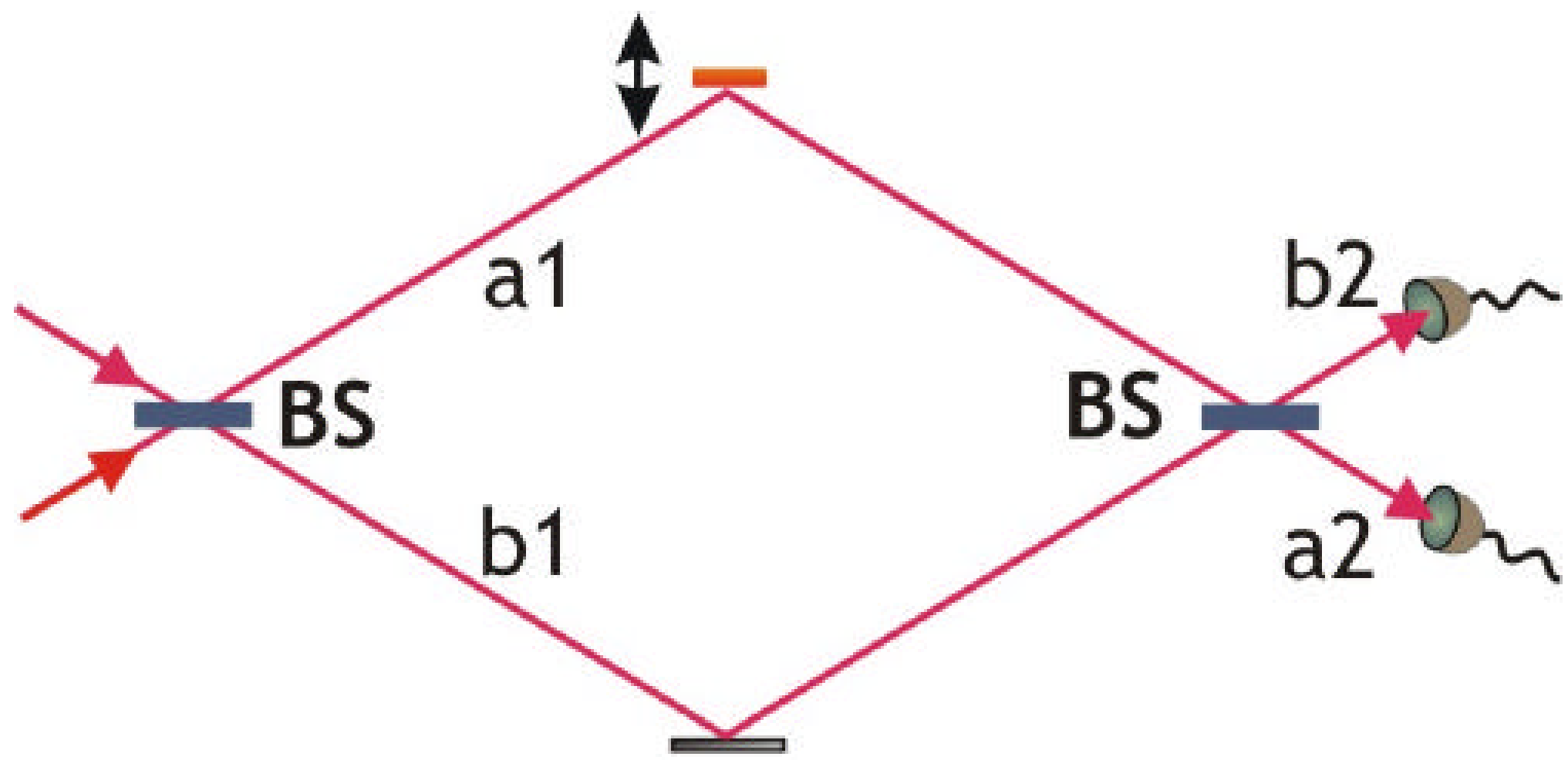

Figure 1

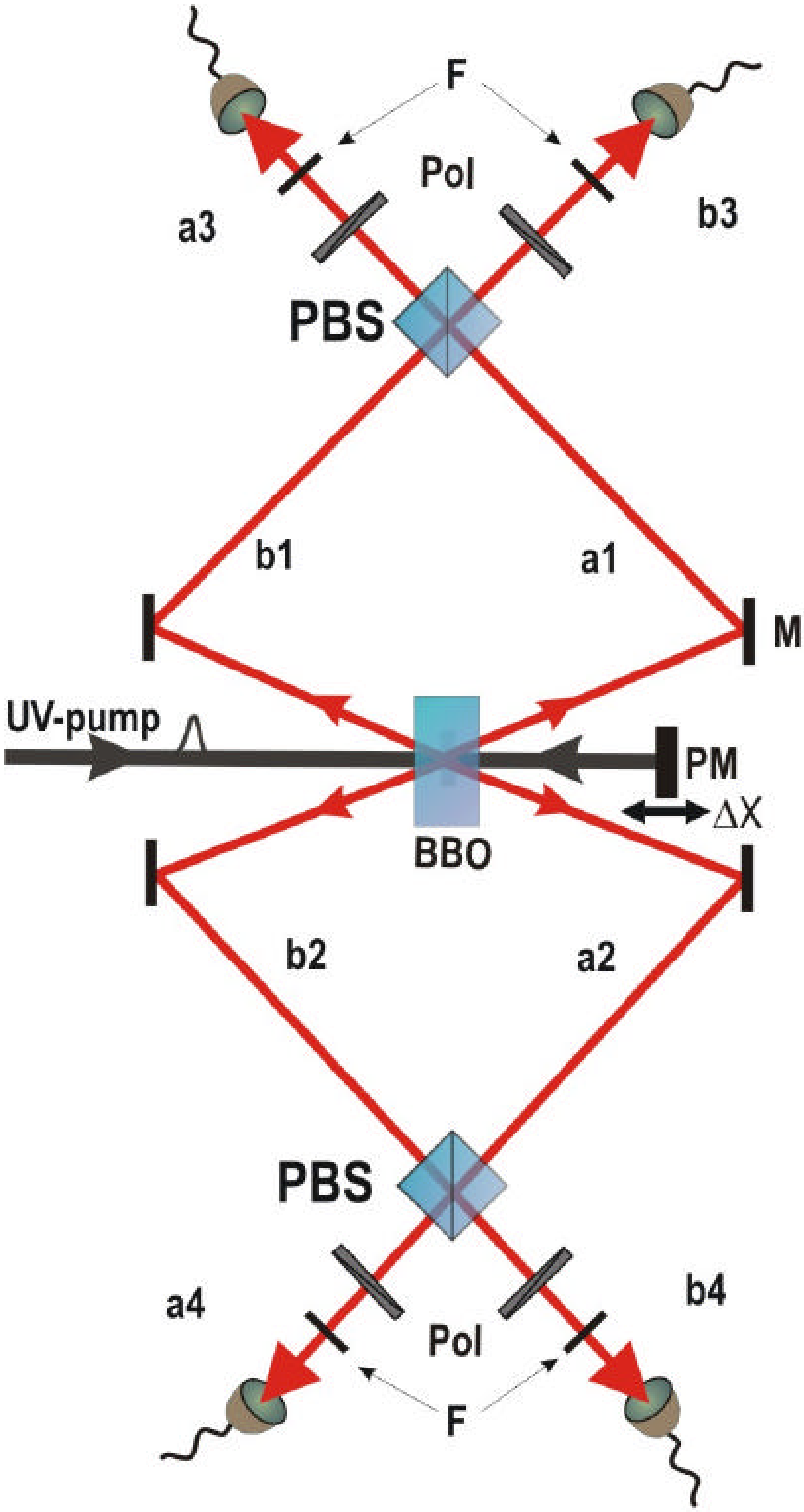

Figure 2

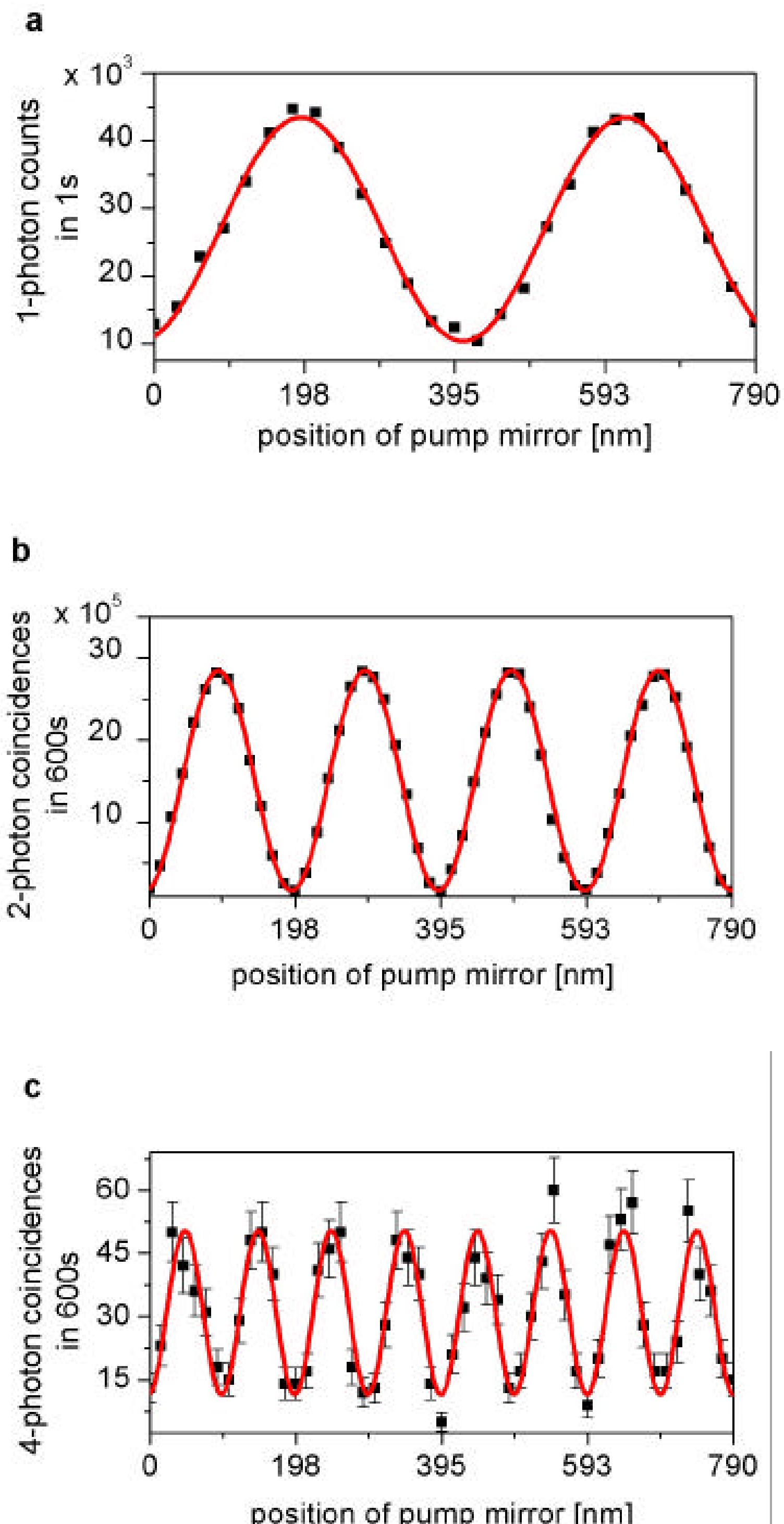
a
x 10³
1-photon counts in 1s
10
20
30
40
0
198
395
593
790
position of pump mirror [nm]
b
x 10⁵
2-photon coincidences in 600s
10
20
30
0
198
395
593
790
position of pump mirror [nm]
c
4-photon coincidences in 600s
15
30
45
60
0
198
395
593
790
position of pump mirror [nm]

Figure 3

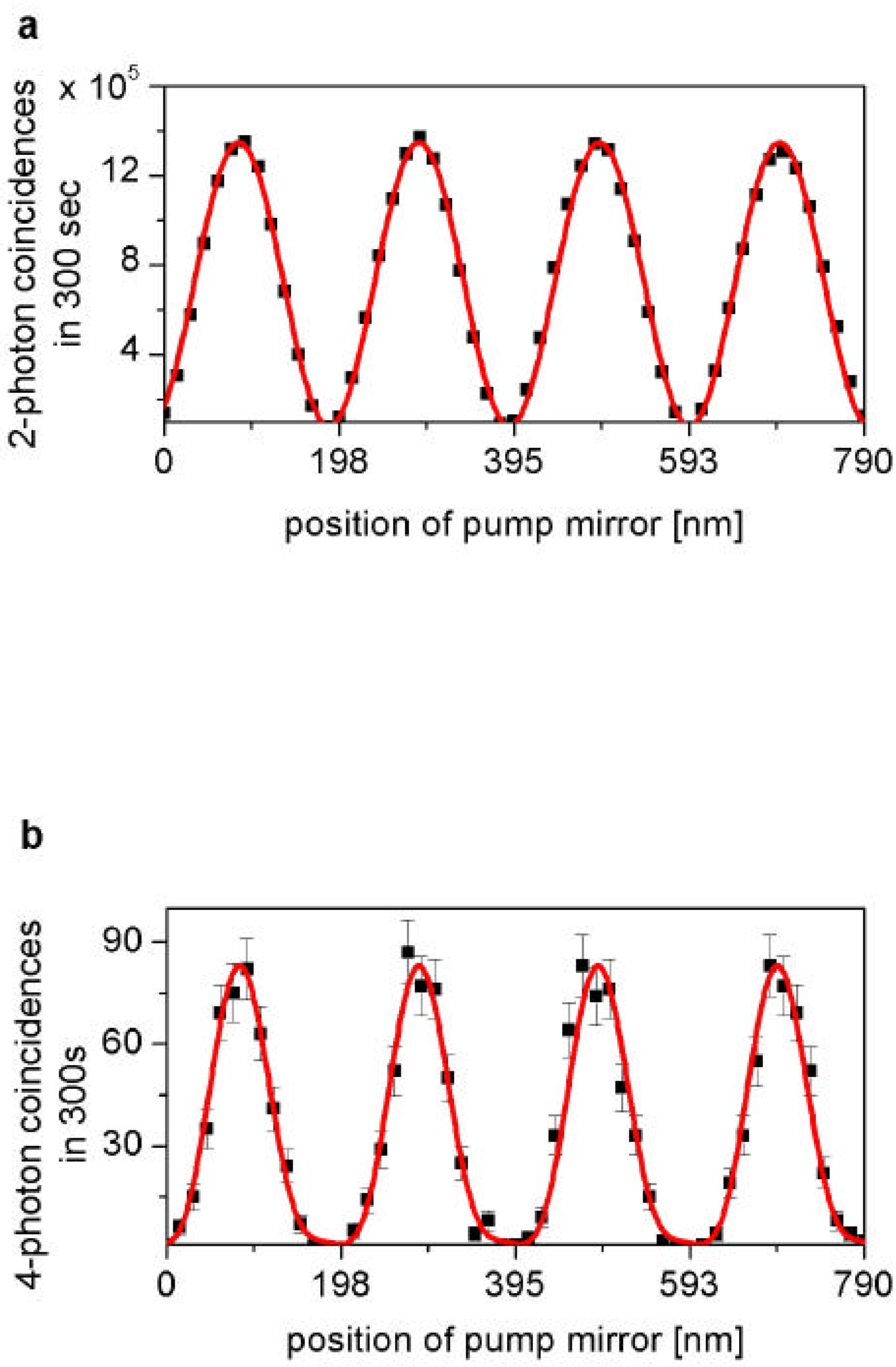
a
2-photon coincidences in 300 sec
x 10^5
12
8
4
0
198
395
593
790
position of pump mirror [nm]
b
4-photon coincidences in 300s
90
60
30
0
198
395
593
790
position of pump mirror [nm]

Figure 4